\documentclass{svjour3}
\usepackage{amsmath}
\usepackage{amssymb}
\usepackage{graphicx}

\usepackage{txfonts}

\usepackage[colorlinks=true,linkcolor=blue,citecolor=blue,urlcolor=blue]{hyperref}

\begin{document}

\title{Encrypted quantum correlations: Delayed choice of quantum statistics and other applications}


\author{Manuel Gessner \and Augusto Smerzi}
\institute{Manuel Gessner \at D\'{e}partement de Physique, \'{E}cole Normale Sup\'{e}rieure, PSL Universit\'{e}, CNRS,
24 Rue Lhomond, 75005 Paris, France \email{manuel.gessner@ens.fr} \and Manuel Gessner \at Laboratoire Kastler Brossel, ENS-PSL, CNRS, Sorbonne Universit\'{e}, Coll\`{e}ge de France, 24 Rue Lhomond, 75005 Paris, France \and Augusto Smerzi \at QSTAR, CNR-INO and LENS, Largo Enrico Fermi 2, 50125 Firenze, Italy}
\date{\today}

\maketitle

\begin{abstract}
In a three-particle extension of Wheeler's delayed choice gedanken experiment, the quantum statistics of two particles is undetermined until a third particle is measured. As a function of the measurement result, the particles behave either as bosons or as fermions. The particles are distinguishable if no measurement is performed at all or when the measurement is performed in a rotated basis. The scheme is based on Greenberger-Horne-Zeilinger quantum correlations. It can be interpreted more generally as the encryption of maximally entangled states in a larger quantum superposition. The local quantum information is scrambled but can be decoded by the measurement result of a control particle. This can be extended to multiple particles and allows to develop quantum information protocols whose successful implementation depends on the collaboration of all parties.
\end{abstract}

\section{Introduction}
Particle-like or wave-like behavior of photons can be observed in an interferometer depending on how the measurement is performed \cite{ScullyBook,Englert}. Wheeler famously raised the question~\cite{Wheeler} of when does the photon ``decide" whether it is going to behave as a wave or as a particle? His gedanken experiments, which today have become an experimental reality \cite{Zeilinger}, lead to the conclusion that the properties of quantum objects are intrinsically undetermined until a measurement is performed. These ideas have been extended to two entangled photons, such that the quantum properties of one of them are conditioned on the measurement results of the other \cite{Scully}. In this case, the wave-like or particle-like nature of one photon can be determined after it was measured, through the delayed measurement of the second photon, which either erases the ``which-way'' information on the first photon or preserves it \cite{Kim}. Again, the quantum properties are not determined until the full system, comprising both photons in an entangled state, has been measured. Since their early conception~\cite{Weizsaecker,Wheeler}, ``delayed-choice'' techniques have been extended to scenarios beyond the wave-particle duality \cite{Peres,Zeilinger} and continue to attract the interest of researchers to this date \cite{Chaves}.

In this article we point out that ``delayed-choice'' schemes can be used more generally to encrypt coherent quantum information of multiparticle quantum states by means of an additional control particle. The quantum properties of the many-particle system are undetermined until the measurement of the control particle is performed. This measurement can be realized in combination with a ``delayed choice" of whether to observe or erase these quantum properties. In the latter case, or in complete absence of this additional measurement, the recorded data is completely scrambled. If a quantum eraser measurement is chosen for the control particle, its measurement result provides the key to unlocking the recorded data and to reveal the quantum properties concealed by quantum randomness.

We illustrate our findings first for the case of a three-particle scenario, where we show that the quantum statistical behavior of two particles can be conditioned onto the measurements of a third one. A modified measurement setup reveals the conditional nonlocality between the two particles as a function of the third one. Finally, we discuss an extension of the scheme to multiple particles based on Greenberger-Horne-Zeilinger (GHZ) states.


\begin{figure}
\centering
\includegraphics[width=.7\textwidth]{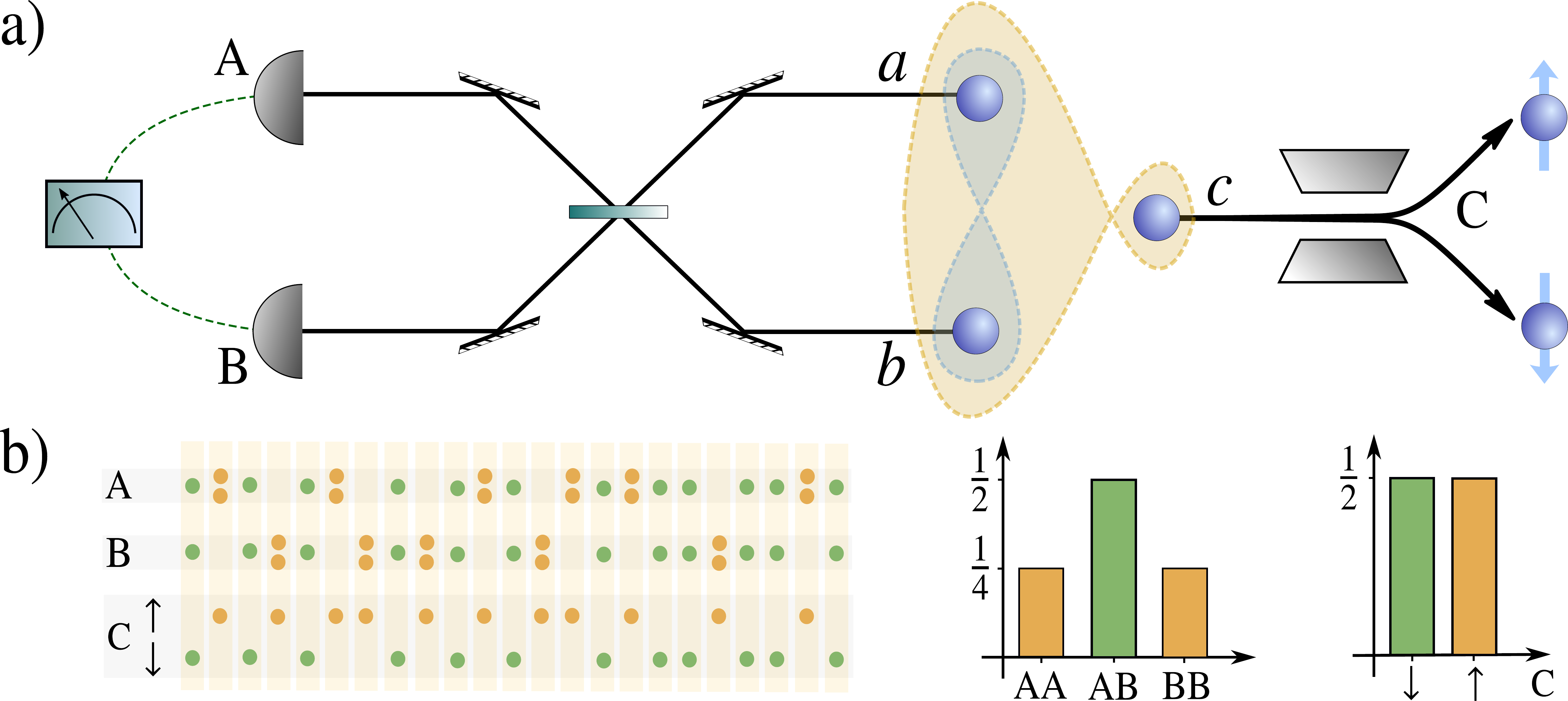}
\caption{a) A Hong-Ou-Mandel experiment on particles $a$ and $b$ probes their quantum statistics, for the initial three-particle quantum state described in Eq.~(\ref{eq:psi}). The measurement result of the spin of particle $c$ determines the quantum statistics of particles $a$ and $b$. b) For $\phi'=0$, bosonic bunching events occur in coincidence with spin $\uparrow$, while fermionic anti-bunching events occur together with the result spin $\downarrow$; see Eq.~(\ref{eq:eventsphi}). The spin state and the quantum statistics randomly take on different outcomes and are both undetermined until all parties are measured.}
\label{fig:1}
\end{figure}

\section{Results and Discussion}
\textit{Delayed choice of quantum statistics.---}We consider three particles (see Fig.~\ref{fig:1}a) with internal (spin-$1/2$) and external degrees of freedom (D.O.F). Let the system be prepared initially in the state
\begin{align}\label{eq:psi}
|\Psi_{\phi}\rangle=\frac{1}{\sqrt{2}}(a^{\dagger}_{\uparrow}b^{\dagger}_{\downarrow}c^{\dagger}_{\rightarrow}+e^{i\phi}a^{\dagger}_{\downarrow}b^{\dagger}_{\uparrow}c^{\dagger}_{\leftarrow})|0\rangle,
\end{align}
where the subscripts $\uparrow$ and $\downarrow$ denote eigenstates of the Pauli matrix $\sigma_z$ with eigenvalues $1$ and $-1$, respectively, and analogously $\rightarrow$ and $\leftarrow$ for $\sigma_x$.
A Hong-Ou-Mandel experiment \cite{HOM} is performed on particles A and B to analyze their quantum statistical behavior. Fermionic behavior is manifested by anti-bunching events, i.e., the two particles never exit the same output port. In contrast, bosonic behavior is reflected by bunching events with both particles always exiting together from the same port. 
The particles are thus submitted to a beam splitter which acts only on the external D.O.F while leaving the spin unchanged. The transformation is described by
$\begin{pmatrix}
a'_{\varphi}\\
b'_{\varphi}
\end{pmatrix}=
\frac{1}{\sqrt{2}}
\begin{pmatrix}
1 & i\\
i &  1
\end{pmatrix}
\begin{pmatrix}
a_{\varphi} \\
b_{\varphi}
\end{pmatrix}$,
for arbitrary spin states $\varphi$. A measurement of the output modes $A$ and $B$ reveals the quantum statistical behavior of the particles in the initial modes $a$ and $b$. Furthermore, the spin of particle $C$ is measured in the $\uparrow/\downarrow$ basis. Assuming the particles to be bosons and for $\phi=0$, we find that the particles exit randomly at the output ports $A$ and $B$; bunching and anti-bunching events occur with equal probability. However, if the results are labeled by the result of the spin of particle $C$, one observes that bunching events are always correlated with a measurement result $\uparrow$ at $C$, while anti-bunching occurs in unison with the result $\downarrow$. 

This can be intuitively understood by rewriting the state~(\ref{eq:psi}) as
\begin{align}\label{eq:spinpsi}
|\Psi_{\phi}\rangle=\frac{1}{\sqrt{2}}(|\Psi_+\rangle\otimes|\uparrow\rangle+|\Psi_-\rangle\otimes|\downarrow\rangle),
\end{align}
with $|\Psi_{\pm}\rangle=(|\uparrow\downarrow\rangle\pm e^{i\phi}|\downarrow\uparrow\rangle)/\sqrt{2}$. Hence, for $\phi=0$, the relative state conditioned on the result $\uparrow$ at $C$ is given by a symmetric spin state for particles $a$ and $b$. Consequently, the external D.O.F of particles $a$ and $b$ must also be symmetric in order to respect the symmetry of the full quantum state of any bosonic many-body system, which leads to the characteristic bunching events. The converse occurs for the measurement result $\downarrow$, where the particles' external D.O.F needs to be anti-symmetric to yield a symmetric total quantum state and demonstrate the anti-bunching, characteristic of fermionic statistics.

These observations can be generalized to arbitrary phases $\phi$, leading to the detection probabilities \cite{probabs}
\begin{center}
\begin{align}\label{eq:eventsphi}
\begin{array}{c|cccc}
& C=\;\uparrow & C=\;\downarrow & C=\;?\\\hline
AB & \frac{1}{2}\sin^2(\phi'/2) & \frac{1}{2}\cos^2(\phi'/2) & \frac{1}{2}\\
AA & \frac{1}{4}\cos^2(\phi'/2) & \frac{1}{4}\sin^2(\phi'/2) & \frac{1}{4}\\
BB & \frac{1}{4}\cos^2(\phi'/2) & \frac{1}{4}\sin^2(\phi'/2) & \frac{1}{4}
\end{array},
\end{align}
\end{center}
where $\phi'=\phi$ for bosons and $\phi'=\phi+\pi$ for fermions. By averaging over the results of $C$, we obtain flat distributions of distinguishable particles, independently of $\phi$ or the nature of the particles.

The measurements on $A$ and $B$ can be collected long before the measurement of $C$ is realized. In this case, one obtains the average distribution which is that of distinguishable particles. Only after also $C$ is measured, the particle statistics for $A$ and $B$ will be determined. By labeling the events according to the measurement result obtained by $C$, the events recorded for $A$ and $B$ can be separated into two different sets whose outcomes are predicted by~(\ref{eq:eventsphi}), cf. Fig.~\ref{fig:1}.

The particles behave as distinguishable not only if $C$ is not measured at all, but also if $C$ is measured in the basis $\rightarrow/\leftarrow$. In this case, the particles in $a$ and $b$ are uniquely labeled by their orthogonal spin states for each of the two measurement results on $C$, as can be seen from Eq.~(\ref{eq:psi}). This corresponds to the analog of a ``which-way'' measurement in an interferometer \cite{Englert}. In contrast, a measurement in the $\uparrow/\downarrow$ basis realizes a quantum eraser measurement of this ``which-way'' information. Setting the measurement basis for $C$ long after the data collection on $A$ and $B$ realizes a quantum ``delayed-choice'' experiment on the quantum statistics.


It is worth emphasizing that the physical particles are at all times either bosons or fermions. Their quantum statistical behavior is manipulated by changing the symmetry of the full three-particle spin wave function. Since the total wave function must obey either bosonic or fermionic statistics, these spin manipulations affect the symmetry of the external D.O.F which is made visible by the Hong-Ou-Mandel experiment. 

\textit{Generalization and applications.---}More generally, we can interpret the considered scenario as the projection of two particles into one of two maximally entangled states with opposite phases, conditioned on the (possibly ``delayed-choice") measurement result performed on the third particle. Without the additional classical information provided by the measurement on $C$ in a suitable basis, the Bell states of $A$ and $B$ described by~(\ref{eq:spinpsi}) are useless for quantum information processing, as their statistical mixture is the separable, zero-discord density matrix $\rho=\frac{1}{2}(|\uparrow\downarrow\rangle\langle\uparrow\downarrow|+|\downarrow\uparrow\rangle\langle\downarrow\uparrow|)$. Indeed, the quantum state~(\ref{eq:psi}) is up to local unitary operations equivalent to a three-particle GHZ state~\cite{GHZ}, $|\rm GHZ\rangle_3=\frac{1}{\sqrt{2}}(|\uparrow\uparrow\uparrow\rangle+|\downarrow\downarrow\downarrow\rangle)$. Its intrinsic symmetry consequently implies that each of the three parties is able to determine the quantum statistical behavior of the other two. In other words, the entanglement and nonlocality among any two parties is encrypted by the third one, and can only be harnessed if all parties collaborate.

This gives rise to a series of applications in quantum information theory, based on the conditional exploitation of two-particle entanglement or nonlocality in combination with a third partner. Any task that relies on Bell states can be modified to a scheme based on GHZ states, such that its successful implementation is possible only if the classical information about the measurement result on $C$ is provided. Specific examples of this kind have already been discussed in the literature. ``Third-man quantum key distribution (QKD)'' \cite{TMC} makes use of the fact that an ideal implementation of Ekert's 1991 scheme for QKD \cite{Ekert} requires a Bell state. In the presence of a state of the class of Eq.~(\ref{eq:psi}), the correlations of the Bell state between $A$ and $B$ can be harnessed only if the measurement results of $C$ are communicated. Hence, $C$ holds the key that is required to establish a successful QKD between $A$ and $B$. Analogously, quantum teleportation from $A$ to $B$ \cite{Teleportation} may be realized with the assistance of $C$ if a state of the type~(\ref{eq:psi}) is initially prepared. A similar scheme based on continous variables has been discussed in Ref.~\cite{vanLoock}. GHZ-type correlations are also at the core of ``quantum secret sharing'' \cite{Hillery}, where a secret is split among three parties such that neither of them can acquire knowledge of it without the collaboration of all the others.

\begin{figure}
\centering
\includegraphics[width=.7\textwidth]{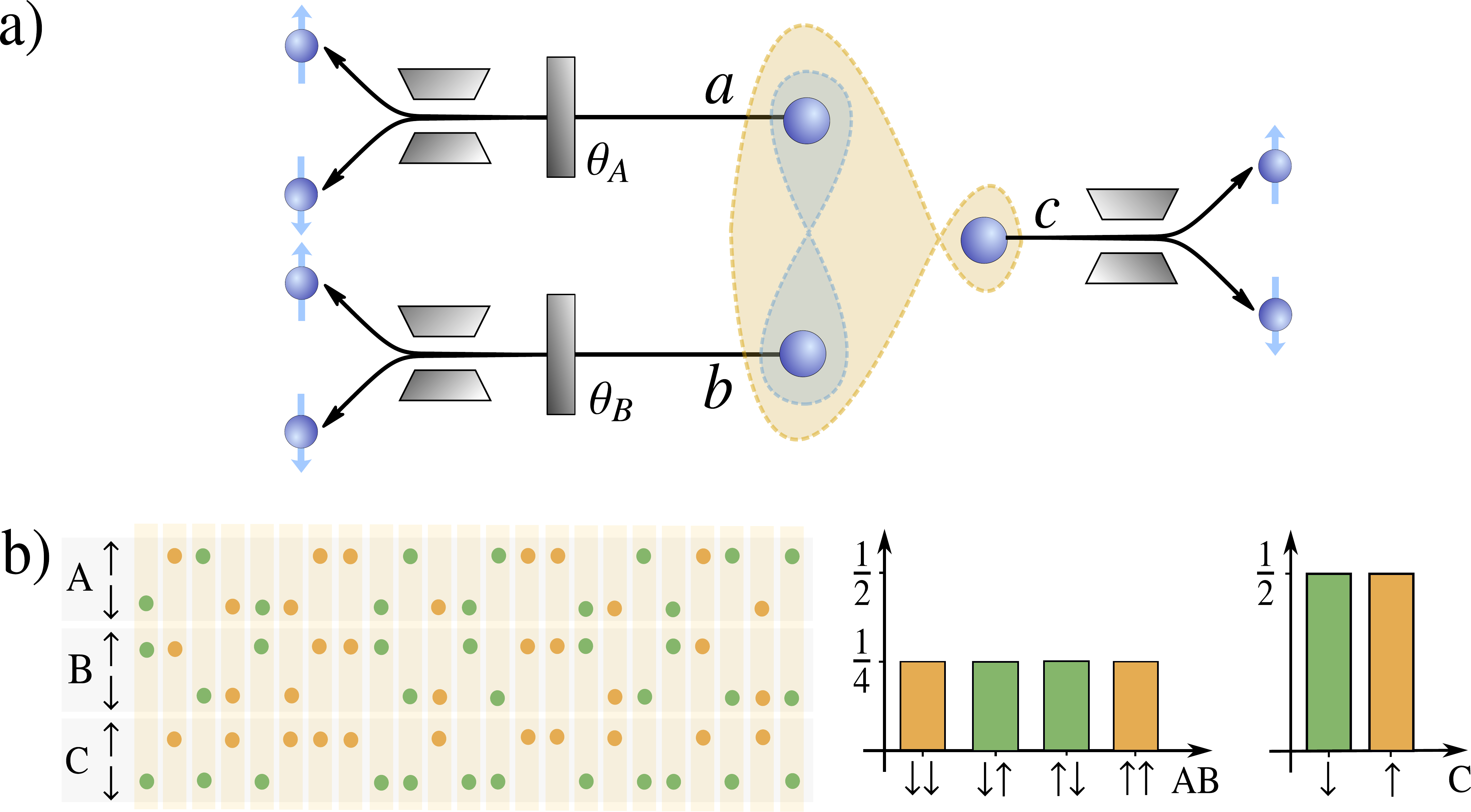}
\caption{a) Local spin measurements on $A$ and $B$ are used to test for nonlocality via a violation of the CHSH inequality, using the quantum state described in Eq.~(\ref{eq:psi}). Only if the measurement result of particle $C$ is known, the data set violates Bell's notion of locality. b) Example of a possible measurement sequence, according to~(\ref{eq:eventsphiCHSH}) for $\phi'=0$. In each subsystem, all events occur randomly. Ordering the events in $AB$ according to the result obtained by $C$ and measuring different combinations of local detector settings reveals the nonlocal nature of the acquired data.}
\label{fig:2}
\end{figure}

\textit{Conditional nonlocality.---}To formally observe the conditional nonlocality in the relative state~(\ref{eq:spinpsi}), we replace the Hong-Ou-Mandel interference setup for $A$ and $B$ by an apparatus that performs local spin measurements in a basis that can be defined by the local angles $\theta_A$ and $\theta_B$, respectively; see Fig.~\ref{fig:2}. A formal verification of the conditional nonlocality of $A$ and $B$ is provided by an analysis based on the Clauser-Horne-Shimony-Holt (CHSH) inequality \cite{CHSH}. A measurement of $\sigma_{\theta_A}\otimes \sigma_{\theta_B}$, with $\sigma_{\theta_i}=\cos(\theta_i)\sigma_x+\sin(\theta_i)\sigma_y$, on particles $A$ and $B$ yields the results $\uparrow$ or $\downarrow$ for $A$ and $B$, according to the probability distribution
\begin{center}
\begin{align}\label{eq:eventsphiCHSH}
\begin{array}{c|cccc}
AB & C=\;\uparrow & C=\;\downarrow & C=\;?\\\hline
\downarrow\downarrow & \frac{1}{4}\cos^2(\phi'/2) & \frac{1}{4}\sin^2(\phi'/2) & \frac{1}{4}\\
\downarrow\uparrow & \frac{1}{4}\sin^2(\phi'/2) & \frac{1}{4}\cos^2(\phi'/2) & \frac{1}{4}\\
\uparrow\downarrow & \frac{1}{4}\sin^2(\phi'/2) & \frac{1}{4}\cos^2(\phi'/2) & \frac{1}{4}\\
\uparrow\uparrow & \frac{1}{4}\cos^2(\phi'/2) & \frac{1}{4}\sin^2(\phi'/2) & \frac{1}{4}
\end{array},
\end{align}
\end{center}
where $\phi'=\theta_A-\theta_B+\phi$. As in Eq.~(\ref{eq:eventsphi}), the distributions are conditioned on the outcome of the measurement at $C$. The conditional expectation values of $\sigma_{\theta_A}\otimes \sigma_{\theta_B}$ are described by the relative states $|\Psi_{\pm}\rangle$ contained in Eq.~(\ref{eq:spinpsi}), multiplied by a factor $1/2$ representing the probability to obtain either of the two results for $C$. We obtain $\langle \sigma_{\theta_A}\otimes \sigma_{\theta_B}\rangle_{|\Psi_{\pm}\rangle}=\pm \cos(\theta_A-\theta_B+\phi)$, where according to~(\ref{eq:spinpsi}), $|\Psi_+\rangle$ is found for $C=\:\uparrow$ and $|\Psi_-\rangle$ occurs together with $C=\:\downarrow$. For each of the two post-selected ensembles, a suitably chosen set $(\theta_A^0,\theta_A^1,\theta_B^0,\theta_B^1)$ of angles \cite{footoptangles} can lead to a maximal violation of the condition $S=|\langle \sigma_{\theta_A^0}\otimes \sigma_{\theta_B^0}\rangle+\langle \sigma_{\theta_A^0}\otimes \sigma_{\theta_B^1}\rangle+\langle \sigma_{\theta_A^1}\otimes \sigma_{\theta_B^0}\rangle-\langle \sigma_{\theta_A^1}\otimes \sigma_{\theta_B^1}\rangle|\leq 2$ up to the value of $2\sqrt{2}$ \cite{Cirelson}, indicating nonlocality \cite{CHSH}. However, if the result of $C$ is not known, a violation of the CHSH inequality is not possible, as $A$ and $B$ do not know which of the two states describes their experiment and their mixture does not feature any quantum correlations.

\textit{Extensions to multipartite settings.---}The principle of the generalized scheme can be extended to multiparty settings, by successively including additional entangled parties whose collaboration is required to make use of the entanglement of the existing set of particles. Such a hierarchy is naturally provided by $N$-particle GHZ states $|\rm GHZ_{\phi}\rangle_N=\frac{1}{\sqrt{2}}(|\uparrow\cdots\uparrow\rangle+e^{i\phi}|\downarrow\cdots\downarrow\rangle)$. Using the equivalent expression
\begin{align}\label{eq:GHZdecomposition}
|\rm GHZ_{\phi}\rangle_{N+1}=\frac{1}{\sqrt{2}}\left(|\rm GHZ_{\phi}\rangle_{N}\otimes|\rightarrow\rangle+|\rm GHZ_{\phi+\pi}\rangle_{N}\otimes|\leftarrow\rangle\right)
\end{align}
reveals the presence of conditional $N$-partite GHZ states, that can be encrypted by one additional party. A measurement of the $(N+1)$th party in the eigenbasis of $\sigma_x$ yields two conditional GHZ states with opposite phases for the remaining $N$ parties. 

The expression~(\ref{eq:GHZdecomposition}) clearly illustrates why each single particle provides the key to the quantum information that is available to the remaining particles, in agreement with what we observed by the three-particle examples discussed before. Without knowledge of the measurement result for the $(N+1)$th particle in a suitably chosen basis, the quantum information of the other $N$ parties remains encrypted, and any measurement will yield scrambled data. As it is fundamentally unknowable which of the two GHZ states is provided, the data is encrypted by genuine quantum randomness. By post-processing the data according to the result of the control particle, the quantum information can be decoded and the quantum correlations can be used. Each party is able to control the quantum information of the others by performing local measurements, due to the symmetry of the states $|\rm GHZ_{\phi}\rangle_N$ \cite{footnotesymmetry}.

Decoding is only possible by quantum eraser measurements in the $\sigma_x$ basis, whereas ``which-way'' measurements in the $\sigma_z$ basis would irreversibly destroy the quantum properties, as the system is projected into one of two product states. The basis can be determined by ``delayed-choice''. The following example of a multiparameter interferometer illustrates these statements.

\begin{figure}
\centering
\includegraphics[width=.7\textwidth]{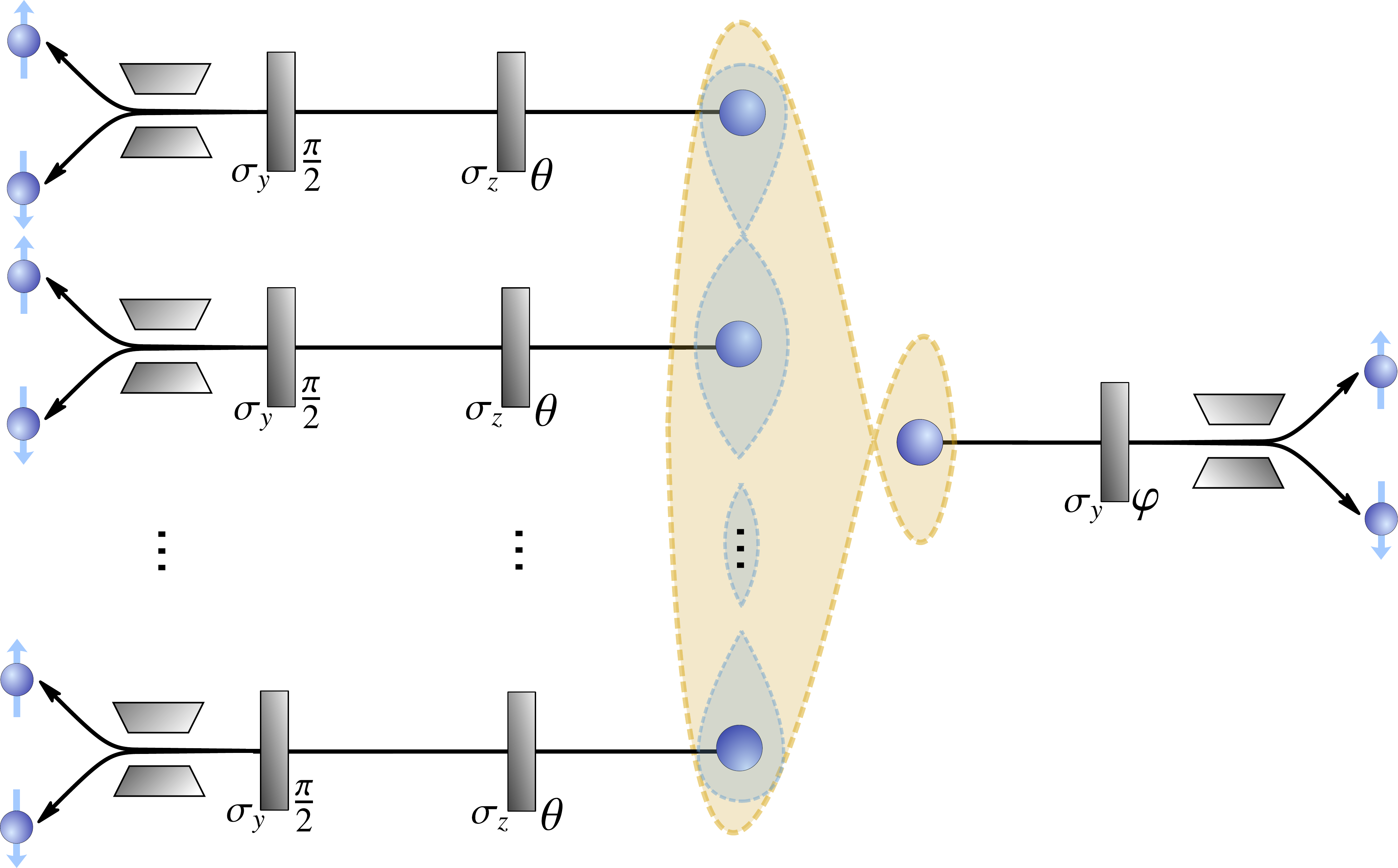}
\caption{In a multipartite extension of the scheme, the sensitivity of an $N$-particle interferometer is dependent on the information of the measurement result of the control particle. The scrambled interferometer data can be decoded by the measurement result of the control particle when $\varphi=\pi/2$. Setting $\varphi=0$ provides ``which-way'' information about the interferometer, making the scrambling of the acquired data irreversible. The angle $\varphi$ can be modified by ``delayed-choice'' after the interferometer measurement has been realized.}
\label{fig:3}
\end{figure}

\textit{``Delayed-choice'' quantum phase estimation.---}GHZ states are the unique family of states that reach the maximum phase sensitivity in an interferometers, commonly known as the Heisenberg limit $(\Delta\theta_{\rm HL})^2=N^{-2}$ \cite{Bollinger,Giovannetti,Varenna}. Consider an $N$-particle interferometer that is entangled with a single additional control particle to form the state $|\rm GHZ_{\phi}\rangle_{N+1}$, see Fig.~\ref{fig:3}. The $N$ particles pass through the interferometer, composed of the collective phase shift $\theta$ and a $\pi/2$-rotation around $y$ that describes a beam splitter operation (left side). A measurement of the parity $P=(-1)^{N/2-J_x}=e^{-iJ_y\pi/2}(-1)^{N/2-J_z}e^{iJ_y\pi/2}$ with $J_{\alpha}=\sum_{i=1}^N\sigma^{(i)}_{\alpha}/2$ for $\alpha=x,y,z$, is realized by measuring $\sigma_z^{(i)}$ for each particle $i=1,\dots,N$ at the end of the sequence. The control particle (right side) is rotated by $e^{-i\sigma_y\varphi/2}$ and subsequently measured in the eigenbasis of $\sigma_z$.

If the control particle is measured with $\varphi=\pi/2$, the results can be used to label the recorded interferometer data into two sets, according to the measurement result (up or down). By virtue of Eq.~(\ref{eq:GHZdecomposition}), a measurement of the control particle in the $\sigma_x$-basis randomly projects the interferometer state into one of two GHZ states with opposite phases. These relative states have parity expectation values of $\langle \mathrm{GHZ}_{\phi}|P| \mathrm{GHZ}_{\phi}\rangle=(-1)^N\cos(N\theta+\phi)$ and $\langle \mathrm{GHZ}_{\phi+\pi}|P| \mathrm{GHZ}_{\phi+\pi}\rangle=(-1)^{N+1}\cos(N\theta+\phi)$, respectively. If the data sets are analyzed separately, the $N$-fold enhanced dependence on $\theta$ gives rise to a phase sensitivity at the Heisenberg limit \cite{Bollinger}, that is achievable via the method of moments \cite{Varenna,GessnerPRL2019}. Without any information about the measurement of the control particle, the measurement results of $P$ average to zero and therefore do not provide any useful information about the value of the phase parameter $\theta$.

If $\varphi=0$, the measurement of the control particle is realized in the basis of $\sigma_z$. As this measurement provides ``which-way'' information about the particles in the interferometer \cite{Englert}, the phase-dependent interference signal is destroyed by this measurement and no phase sensitivity is provided. In order to access the highly precise information about the parameter $\theta$ that is contained in the interferometer data, it must be decoded by providing the key that is the measurement result of the control particle for $\varphi=\pi/2$. Choosing the angle $\varphi$ thus allows us to obtain ``which-way'' information that destroys the interferometer signal or to perform a quantum erasure measurement on it and thereby decode the scrambled interferometer data in post-processing. As the measurement of the control particle can be realized long after the interferometer data has been acquired, controlling the rotation angle $\varphi$ enables us to realize a ``delayed-choice'' quantum phase estimation experiment.

The fact that any of the particles in a GHZ state can play the role of the control particle illustrates why Heisenberg-limited quantum phase estimation with GHZ states is so intrinsically fragile \cite{HuelgaPRL1997,Rafal}: There is no tolerance to particle loss and even measurements of a single particle in a wrong basis would destroy the signal.



\textit{Role of quantum correlations.---}Our scheme is based on the property of GHZ states that the absence of information about one particle leaves the remaining ones in a classical mixture of two GHZ states with no quantum correlations. A similar effect is achieved by randomly preparing one of two GHZ states with opposite phases. In this case the control particle (which is quantum) is replaced by a classical agent who holds the information about the random outcome of the state preparation that is not available to the receivers of these states. This situation describes a ``classical encryption'' of the quantum information which is recovered by the more general quantum scheme for the fixed choice of the quantum eraser measurement basis. Hence, classical correlations are sufficient to establish most of the phenomena that were described in this article. However, in contrast to the scheme based on multiparticle GHZ correlations, the system properties are compatible with local realism (hence the encryption is no longer based on quantum randomness) and a quantum ``delayed-choice'' measurement is no longer possible.

\section{Conclusions}
Quantum coherent properties, such as the symmetry of the wave function or the ability to violate Bell's inequality, can be encrypted by an external control particle. This is enabled by the multipartite quantum correlations of GHZ states. Depending on the choice of basis, a measurement of the control particle can either provide the key to decoding the quantum information or irretrievably destroy it. The two possibilities formally correspond to a quantum eraser measurement and the retrieval of ``which-way'' information, respectively. The measurement of the control particle can be realized long after the implementation of some quantum information protocol on the remaining particles, in the spirit of ``delayed-choice'' experiments. We illustrated our results with the quantum statistics of two particles that is conditioned on the measurement of a third particle, the conditional violation of Bell's inequality, and the Heisenberg-limited estimation of a phase parameter. The scrambling and unscrambling of quantum information through an external controller opens up new possibilities to modify quantum information protocols that are based on multipartite entangled GHZ states, such that their implementation is conditioned on the collaboration of the controller.

\section{List of abbreviations}
\begin{itemize}
\item GHZ states: Greenberger-Horne-Zeilinger states.
\item D.O.F: Degree of freedom.
\item QKD: Quantum key distribution.
\item CHSH inequality: Clauser-Horne-Shimony-Holt inequality.
\end{itemize}

\section{Declarations}
\begin{itemize}
\item \textbf{Availability of data and material.}\\
The datasets used and/or analysed during the current study are available from the corresponding author on reasonable request.
\item \textbf{Competing interests.}\\
The authors declare that they have no competing interests.
\item \textbf{Funding.}\\
This work was funded by the LabEx ENS-ICFP:ANR-10-LABX-0010/ANR-10-IDEX-0001-02 PSL*.
\item \textbf{Authors' contributions.}\\
All authors conceived and discussed the idea. M.G. wrote the manuscript.
\item \textbf{Acknowledgements.}\\
Does not apply.
\end{itemize}


\begin{thebibliography}{xx}
\bibitem{probabs} The state $|\Psi\rangle$ is expressed in terms of detector modes $a'^{\dagger}$, $b'$ and $c$ as $|\Psi\rangle=\frac{e^{-i\phi}}{\sqrt{2}}\{i[-a_{\uparrow}'^{\dagger}b_{\downarrow}'^{\dagger}\sin(\phi/2)+b_{\uparrow}'^{\dagger}a_{\downarrow}'^{\dagger}\sin(\phi/2)+b_{\uparrow}'^{\dagger}b_{\downarrow}'^{\dagger}\cos(\phi/2)+a_{\uparrow}'^{\dagger}a_{\downarrow}'^{\dagger}\cos(\phi/2)]c_{\uparrow}^{\dagger}+[a_{\uparrow}'^{\dagger}b_{\downarrow}'^{\dagger}\cos(\phi/2)-b_{\uparrow}'^{\dagger}a_{\downarrow}'^{\dagger}\cos(\phi/2)+b_{\uparrow}'^{\dagger}b_{\downarrow}'^{\dagger}\sin(\phi/2)+a_{\uparrow}'^{\dagger}a_{\downarrow}'^{\dagger}\sin(\phi/2)]c^{\dagger}_{\downarrow}\}|0\rangle$,
where we assumed bosonic particle statistics. For fermions the same expression holds after replacing $\phi$ by $\phi+\pi$. The probabilities for the detection events in A and B are obtained from the above expression by summing over all spin states.
\bibitem{footoptangles} The phase $\phi$ of the superposition can be absorbed into the measurement settings. Splitting $\phi$ symmetrically between $A$ and $B$, a maximal violation \cite{Cirelson} of the CHSH inequality can be achieved, e.g., by $\theta_A^{0}+\phi/2 = 0$, $\theta_A^{1}+\phi/2 = \pi$, $\theta_ B^{0}-\phi/2 = \pi/4$, and $\theta_B^{1}-\phi/2 = -\pi/4$.
\bibitem{footnotesymmetry} The phase $\phi$ can be modified via local rotations on a single party around $\sigma_z$ (local rotations of an arbitrary party around $\sigma_z$ by the angle $\theta$ change the value of the phase $\phi$ to $\phi-\theta$), but its value does not influence the quantum information of the remaining particles. This is different for the state~(\ref{eq:spinpsi}), where the symmetry properties of the conditional states for $A$ and $B$ depend on $\phi$, and thus may lead to bosonic or fermionic phenomenology as a function of $\phi$.
\end{thebibliography}
\end{document}